\documentclass[aps,twocolumn,prl]{revtex4}
\usepackage{color}

\usepackage{epsfig}
\usepackage{epstopdf}
\usepackage{amsmath}
\usepackage{amsfonts}
\usepackage{verbatim}
\usepackage{amssymb}
\usepackage[latin1]{inputenc}
\usepackage{graphicx}
\usepackage[normalem]{ulem}

\newcommand{\nc}{\newcommand}
 \nc{\tcb}{\textcolor{blue}}
 \nc{\tcr}{\textcolor{red}}
 \nc{\be}{\begin{equation}}
 \nc{\ee}{\end{equation}}
 \nc{\bt}{\begin{tabular}}
 \nc{\et}{\end{tabular}}
 \nc{\bea}{\begin{eqnarray}}
 \nc{\eea}{\end{eqnarray}}
 \nc{\ba}{\begin{array}}
 \nc{\ea}{\end{array}}
 \nc{\rds}{{\rm d}s}
 \nc{\rdt}{{\rm d}t}
 \nc{\rdr}{{\rm d}r}
 \nc{\rdO}{{\rm d}\Omega}
 \nc{\s}{{\rm S}}
 \nc{\Pl}{{\rm Planck}}
 \nc{\dis}{\displaystyle}
 \nc{\crit}{_{\rm cr}}
 \nc{\rd}{{\rm d}}
 \nc{\munu}{{\mu\nu}}
 \nc{\erm}{{\rm e}}
 \nc{\drm}{{\rm d}}
 \nc{\ov}{\overline}

\begin{document}

\title{{Phase time and transmission probability in the traversal of a $PT$-symmetric potential: the case of an electromagnetic waveguide}}
\author{M. Di Mauro}
\affiliation{Dipartimento di Fisica ``E.R. Caianiello", Universit\'a di Salerno, Via Giovanni Paolo II, I-84084 Fisciano (SA), Italy}
\author{S, Esposito, A. Naddeo}
\affiliation{Istituto Nazionale di Fisica Nucleare, Sezione di Napoli, Complesso Universitario di Monte
S.\,Angelo, via Cinthia, I-80126 Naples, Italy}

\begin{abstract}
\noindent We study the unconventional transmission properties of a wave-packet through a $PT$-symmetric potential region, as describing actual electromagnetic wave propagation along a waveguide filled with gain and loss media. The non-trivial behavior of the transmission probability manifests in the giant amplification of the incident electromagnetic signal of given wavelengths for well-defined configurations, depending on the gain/loss contrast. Maximum transmission peaks are related to spectral singularities and a strict correlation exists between the ``resonant" wavelengths and the gain/loss contrast. The transit times are as well calculated, showing their surprising vanishing in the opaque barrier limit, independently of the gain/loss contrast, which is reminiscent of some sort of Hartman effect. Also, non-local effects manifest in the presence of negative delay times for given configurations, while a correlation is apparent between maximum delay times and transmission probability peaks, though appreciably depending on the gain/loss contrast.
\end{abstract}

\maketitle

\noindent It has been known for a while that non-Hermitian Hamiltonians enjoying $PT$-symmetry can lead to sensible quantum mechanical models \cite{Bender:1998ke, Bender:1998gh}. Such Hamiltonians, indeed, have real spectra and, upon a proper modification of the inner product in the Hilbert space, exhibit unitary time evolution \cite{Bender:2002vv, Mostafazadeh:2001nr}. This has favored, in recent years, a lot of activity (see e.g. \cite{Bender:2007nj, Bender:2015aja} and references therein), with applications ranging from condensed matter systems \cite{Longhi2009} to quantum field theory \cite{Bender:2004sv} and integrable models \cite{Dorey:2009xa}.

\begin{figure}
\begin{center}
\includegraphics[width=7cm]{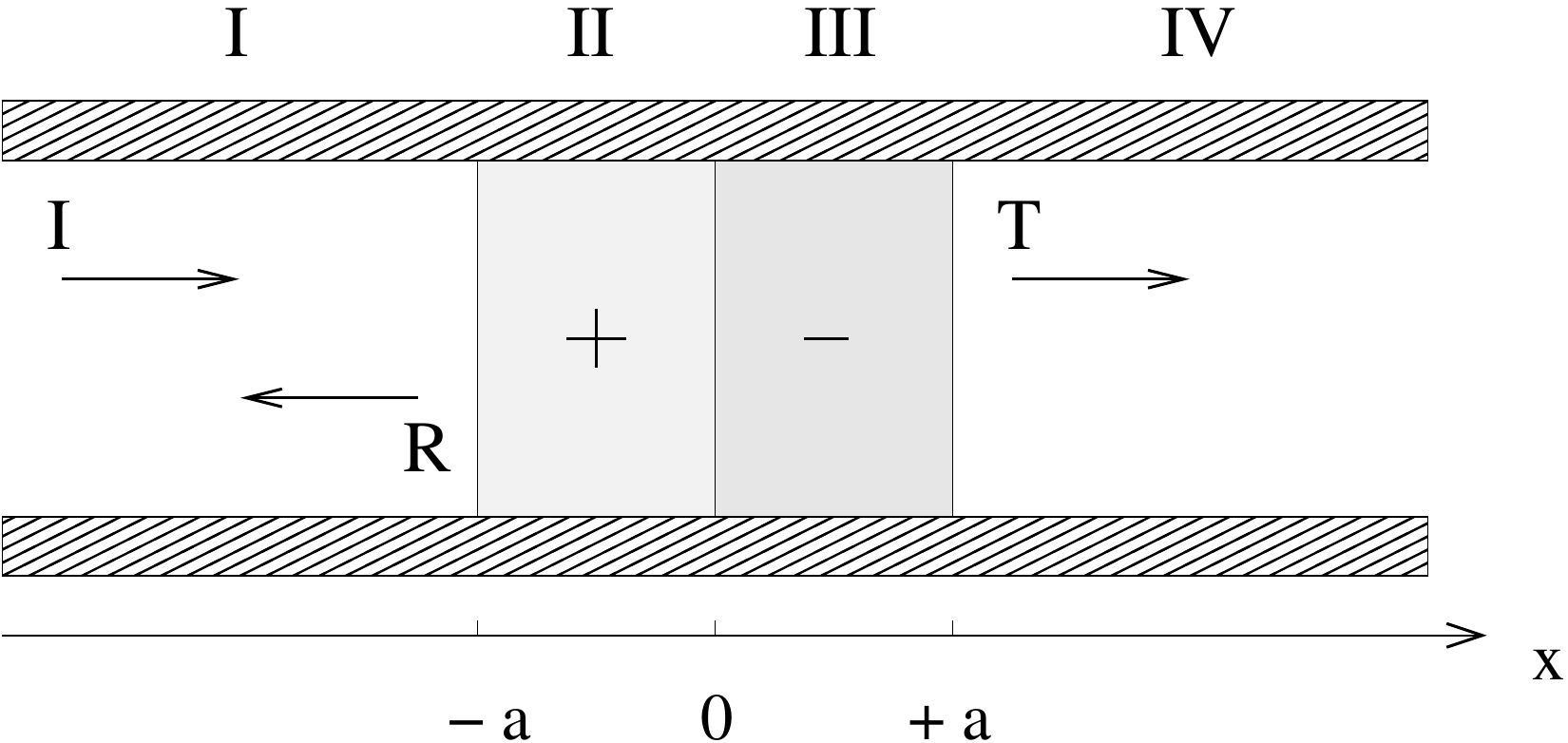}
\caption{Incident (I), reflected (R) and transmitted (T) from a waveguide with gain (+) and loss ($-$) regions, as modelled by the $PT$-symmetric system considered in the paper.}
\label{fig1}
\end{center}
\end{figure}

Particularly interesting is the study of complex $PT$-symmetric scattering potentials {in one \cite{ChongGeStonePRL2011,GeChongStonePRA2012,AmbichlPRX2013} and higher dimensions \cite{GeMakrisChristodoulidesFengPRA2015}. Such systems can be easily realized in optical systems \cite{Muga, ElgananinyOptLett2007,MakrisPRL2008,MusslimaniPRL2008,GuoPRL2009,RueterNature2010}{. Indeed, it has been found that the equation governing the electric field envelope of an optical beam in a waveguide is formally analogous to a Schr\"{o}dinger equation with a suitable potential related to the refraction index or to the permittivity of the media inside the waveguide. The possibility of having gain and loss media allows, thus, to study experimentally in such media the effects of complex potentials and, in particular, of $PT$-symmetric ones, which are constrained to satisfy the condition
\begin{equation}
V(x)=V^*(-x).
\end{equation}
The mentioned optical systems have been widely studied, also from an experimental point of view, and} display a host of interesting phenomena, such as quantum noise induced self-sustained radiation \cite{SchomerusPRL2010}, simultaneous laser-absorbers \cite{LonghiPRA2010} and unidirectional invisibility \cite{LinRamezanietalPRL2011}.}
{In the following we consider} the Schr\"{o}dinger equation associated to the purely imaginary $PT$-symmetric potential \cite{Muga}
\begin{equation}\label{potential}
  V(x)=\begin{cases}
    i \, v,  & \text{for $-a<x<0$,}\\
    -i \, v, & \text{for $0<x<a$,}\\
    0,       & \text{otherwise},
  \end{cases}
\end{equation}
with $a\in \mathbb{R}^+$ and $v\in\mathbb{R}-\{0\}$. {Such an equation} describes electromagnetic waves propagating along a planar slab waveguide (see Fig.~\ref{fig1}) filled with gain and absorbing media, so that a complex permittivity is present for a certain range of the parameters  \cite{Muga}. For example, if the waveguide is filled with an atomic gas, a laser beam impinging on it (perpendicularly to the plane of Fig.~\ref{fig1}) in region II may induce resonant excitation of the atoms, in order to produce a population inversion in such region, while the unaltered region III serves as the loss counterpart. More in general, the real part of permittivity in $PT$-symmetric structures must be an even function of the position vector, while its imaginary (gain/loss profile) must be antisymmetric.
From the mathematical point of view, non-Hermitian Hamiltonians whose spectrum has a continuous part can show mathematical obstructions to the completeness of the eigenvectors, known as ``spectral singularities" \cite{Najmark}. {These spectral singularities correspond to zero-width resonances, i.e. to real values of the energies at which the reflection and transmission coefficients tend to infinity \cite{Mostafazadeh:2009zz, Mostafazadeh:2009ew}. {Notice that the model we study refers to a setup different from the one studied in \cite{ElgananinyOptLett2007,MakrisPRL2008,MusslimaniPRL2008,GuoPRL2009,RueterNature2010} since, in the present case, the permittivity varies along the propagation direction of the wave. This is essential in order to realize the resonance effect related to spectral singularities, as explained in \cite{Mostafazadeh:2009zz}.} Physically, {spectral singularities} correspond, in the optical case, to the lasing thresholds of the gain media and to the coherent perfect absorber (CPA) \cite{ChongGeCaoStonePRL2010} threshold for of the loss media \cite{ChongGeStonePRL2011, MostafazadehPRA2011}. In particular $PT$-symmetric systems can exhibit net optical amplification or absorption at special points \cite{ChongGeStonePRL2011} where $PT$-symmetry is spontaneously broken \footnote{Recall that, if $PT$-symmetry is realized, no net gain or loss can occur.}. Such exceptional points are also those at which $PT$-symmetric optical systems can act as unidirectional invisible media \cite{LinRamezanietalPRL2011}. }

Optical enhancement is, indeed, on the spotlight, and the recent works (see Ref. \cite{Kuwar} and refs. therein) aimed at achieving amplification by using large-scale hybrid quantum systems well justifies further theoretical work to complement the current experimental demonstrations. $PT$-symmetric systems offer, then, a viable theoretical description which still has to display its full potential.

Indeed, if the reflection/transmission through such $PT$-symmetric scattering potentials has been studied to a certain extent, the same does not apply to the problem of the transit times for a wave packet propagating through such regions, including tunneling times for under-barrier configurations. To this regard, it is intriguing that, despite existing different definitions proposed in the literature for the computation of these times (see e.g \cite{Olkhovsky:1991wc, Privitera:1900zz} and references therein), the experimentally favored {\it phase time}
\begin{equation} \label{time}
\tau=\frac{\drm \varphi}{\drm \omega}=\frac{\drm \varphi}{\drm k}\frac{\drm k}{\drm \omega} \, ,
\end{equation}
$\varphi$ being the phase difference acquired by the wave $\erm^{i (k x - \omega t)}$ in traversing the considered region, applies without the need of any modification even in the presence of complex potentials, where evanescent propagation (i.e. tunneling) adds to usual wave propagation.

{In media with gain and loss, the definition of time is even a more delicate issue with respect to the case of non-complex materials, for which we address the interested reader to the mentioned literature. Here, for the sake of physical clarity, we only point out that, in general, the phase time defined above would measure the time delay of the amplitude envelopes of the given components of the electromagnetic wave propagating through the device considered. Such a quantity, however, is physically meaningful only when the output signal is just the result of direct scattering of the input signal from the medium. Nevertheless, in the presence of complex media with gain and loss, the situation is of course at variance with that described, since the output signal has, in general, even components genuinely generated by the medium (or subtracted by the medium), though causally connected to the input signal. In such a case, the phase time defined above is just a very simple useful quantity providing valuable information about the time delay (or advancement) experienced by the transmitted electromagnetic signal when compared to the corresponding one in absence of the scattering medium \cite{Martin}. This is, indeed, exactly what is of interest in experimental investigations, and we definitely refer to the definition above.}

In the present paper we further study in some detail the traversal of a $PT$-symmetric potential barrier, with particular reference also to tunneling times, in order to have a fully physical picture of the effects produced by the presence of spectral singularities.

Our base model is just that generated by the potential in Eq. (\ref{potential}). The continuous spectrum of the Hamiltonian $\dis H=-\frac{\drm^2}{\drm x^2}+  V(x)$ is doubly degenerate \cite{Mostafazadeh:2009zz}, and the left- and right-going scattering solutions of the Schr\"{o}dinger equation $H\psi=k^2\psi$ in regions I and IV can be written as follows:
\begin{eqnarray}\label{leftrightpsi}
\psi_{\rm I}(x) &=& \begin{cases}
    \psi^{\rm I}_{L} = N_L \left( \erm^{i k x} + R_L \, \erm^{-i k x} \right), \\
    \psi^{\rm I}_{R} = N_R \left( T_R \, \erm^{- i k x} \right),
  \end{cases} \,\,(x \rightarrow - \infty) \nonumber \\
  & & \\
\psi_{\rm IV}(x) &=& \begin{cases}
    \psi^{\rm IV}_{L} = N_L \left( T_L \, \erm^{i k x} \right), \\
    \psi^{\rm IV}_{R} = N_R \left( \erm^{-i k x} + R_R \, \erm^{i k x} \right) .
  \end{cases}  \!\!\! (x \rightarrow + \infty) \nonumber
\end{eqnarray}
Here we have denoted with $R,T$ the reflection/transmission amplitudes, while the quantities $N$ are normalization factors. In the active regions II, III both evanescent and oscillating waves are present:
\begin{eqnarray}\label{activepsi}
\psi_{\rm II}(x) &=& \begin{cases}
    \psi^{\rm II}_{L} = A_1 \, \erm^{\delta x} \erm^{i \gamma x} + B_1 \, \erm^{- \delta x} \erm^{- i \gamma x}, \\
    \psi^{\rm II}_{R} = C_1 \, \erm^{- \delta x} \erm^{- i \gamma x} + D_1 \, \erm^{\delta x} \erm^{i \gamma x},
    \end{cases} \nonumber \\
  & & \\
\psi_{\rm III}(x) &=& \begin{cases}
    \psi^{\rm III}_{L} = A_2 \, \erm^{- \delta x} \erm^{i \gamma x} + B_2 \, \erm^{\delta x} \erm^{- i \gamma x}, \\
    \psi^{\rm III}_{R} = C_2 \, \erm^{\delta x} \erm^{- i \gamma x} + D_2 \, \erm^{- \delta x} \erm^{i \gamma x},
    \end{cases} \nonumber
\end{eqnarray}
where the energy (frequency) parameters in the gain ($\omega_+$) and loss ($\omega_-$) regions are defined by \footnote{In general, the dispersion relations in the two active regions II, III need not be equal, though having the same functional form, so that we could expect that $\gamma_{II} \neq \gamma_{III}$ and $\delta_{\rm II} \neq \delta_{\rm III}$. However, for simplicity, we here consider just one couple of parameters for the active system described by the single parameter Eq. (\ref{potential}), as customary in the literature.} $\omega_\pm = \sqrt{E \mp i \, v } = k (\gamma \mp i \, \delta)$, with ($\xi = v^2/E^2$)
\begin{equation} \label{gammadelta}
\gamma = \sqrt{\frac{\sqrt{1+\xi^2} + 1}{2}} \, , \quad
\delta = \sqrt{\frac{\sqrt{1+\xi^2} - 1}{2}} \, .
\end{equation}

\begin{figure*}
\begin{center}
\bt{ll}
\includegraphics[width=16cm]{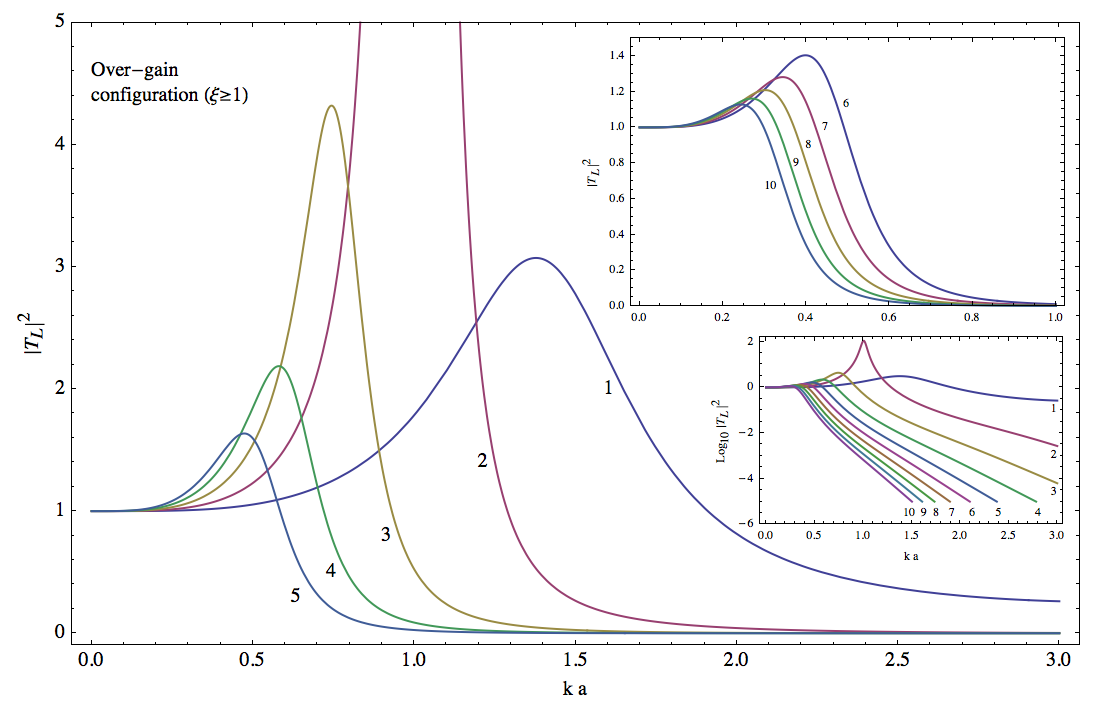}  \\
\includegraphics[width=16cm]{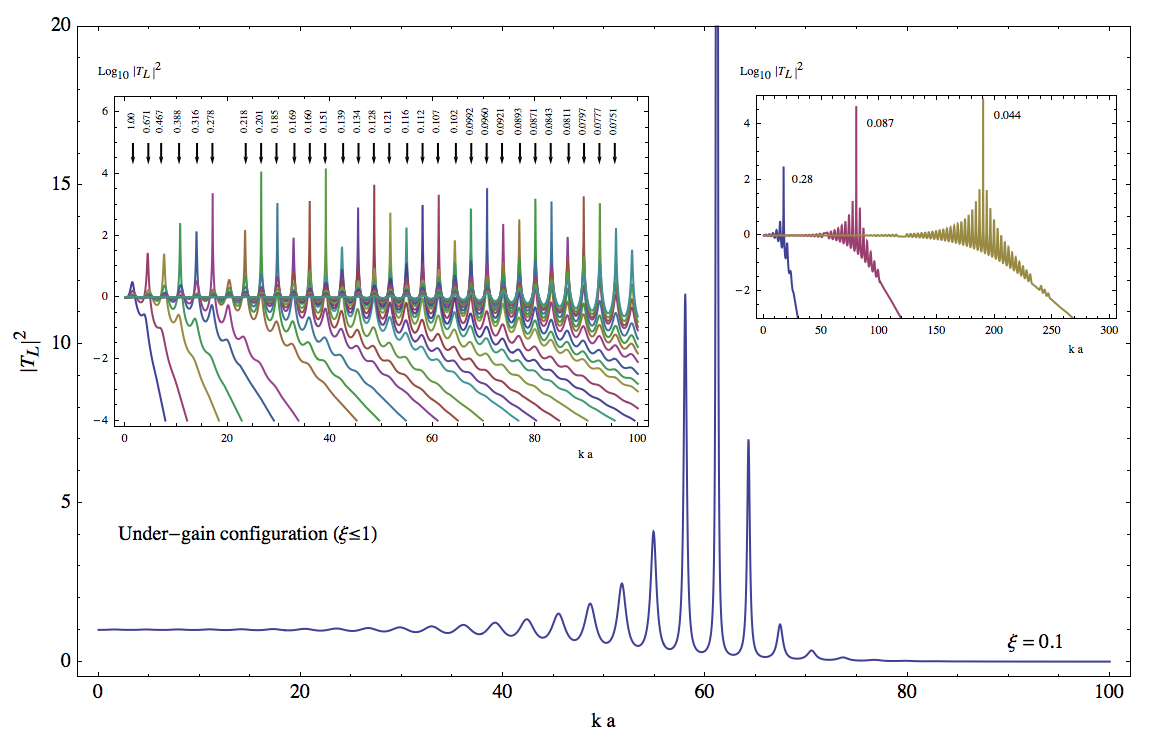}  \et
\caption{Transmission probability for the complex potential system in Fig.~\ref{fig1}, for over-gain (top) and under-gain (bottom) configurations. The numbers appearing in each picture report the values of the $\xi$ parameter considered.}
\label{fig2}
\end{center}
\end{figure*}

\noindent Notice that $\gamma^2=\delta^2+1$, and that the bigger $k$, the more dominant is the propagative contribution, as expected. The various coefficients appearing above may be computed as usual by the matching conditions:
\begin{eqnarray}
\psi_{\rm I} (-a) = \psi_{\rm II} (-a) , \quad & &  \psi_{\rm I}^\prime (-a) = \psi_{\rm II}^\prime (-a) , \nonumber \\
\psi_{\rm II} (0) = \psi_{\rm III} (0) , \quad & &  \psi_{\rm II}^\prime (0) = \psi_{\rm III}^\prime (0) , \nonumber \\
\psi_{\rm III} (a) = \psi_{\rm IV} (a) , \quad & &  \psi_{\rm III}^\prime (a) = \psi_{\rm IV}^\prime (a) . \nonumber
\end{eqnarray}
{Since the transmission coefficients are independent of the incidence direction, $T_L=T_R$ {(this is a general feature of all symmetric complex scattering potentials \cite{Ahmed2001}, and our computations confirm it)},  we can consider a transmission probability given just by:}
\begin{eqnarray}
|T_L|^2  \! &=& \! \left(1-2 \gamma^2\right)^2 \left\{ \left[\gamma^2 \cos (2 \gamma k a) + \delta^2 \cosh (2 \delta  k a) \right]^2 \right. \nonumber \\
& & \left. + \left[\gamma^3 \sin (2 \gamma k a)-\delta^3 \sinh (2 \delta  k a)\right]^2 \right\}^{-1} ,
\label{prob}
\end{eqnarray}
depending only on the two dimensionless quantities $\xi$ and $k a$. {We have verified that the transmittance $|T_L|^2$ and the left and right reflectances $|R_{L,R}|^2$ satisfy the generalized unitarity relation \cite{GeChongStonePRA2012, MostafazadehJPhysA2014,GeMakrisChristodoulidesFengPRA2015}.
{\begin{eqnarray}
|T_L|^2-1 = |R_L R_R|,
\end{eqnarray}}
as is appropriate for a $PT$-symmetric scattering potential.}

\begin{figure*}
\begin{center}
\bt{ll}
\includegraphics[width=12cm]{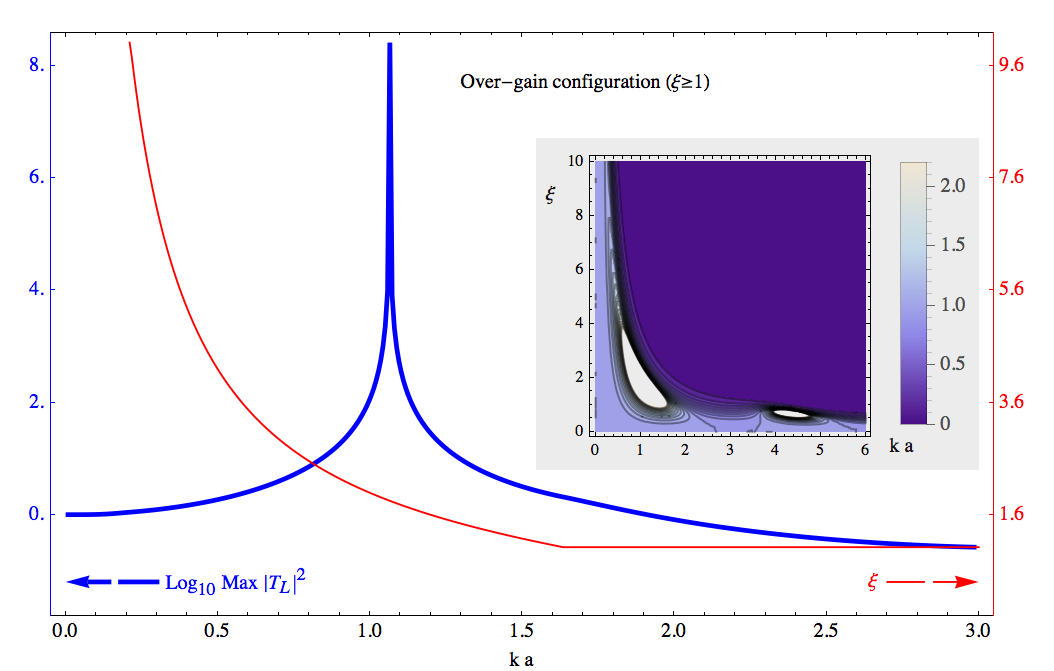}  \\
\includegraphics[width=12cm]{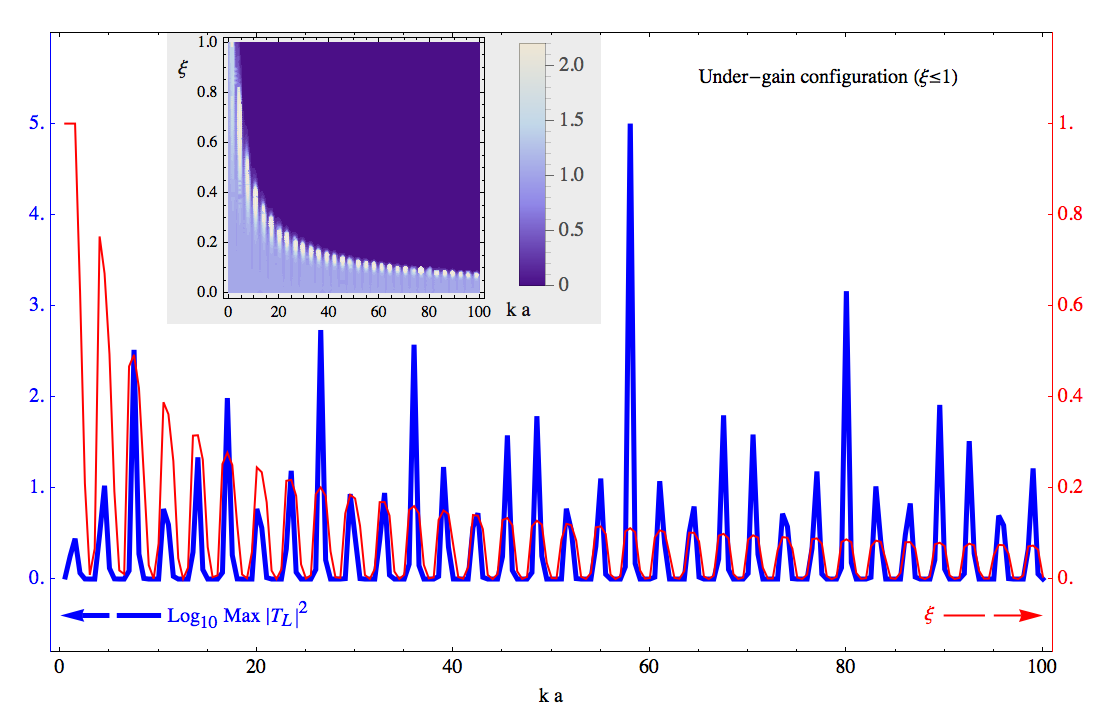}  \et
\caption{Distribution of the transmission probability maxima with $k a$ (thick, blue lines; left scale), and correlation of the position in $k a$ of such maxima with the ``resonant" $\xi$ values (thin, red lines; right scale). In particular, for the over-gain configuration (top), the distribution of the only one peak height is showed; while, for the under-gain configuration (bottom), the distribution of just the higher probability maximum (maximum of the probability wave packet in Fig.~\ref{fig2} bottom) is showed. The inserts report the contour plots of the transmission probability for the two configurations (darker regions refer to lower probabilities).}
\label{fig3}
\end{center}
\end{figure*}

The study of the spectral singularities, which materialize when the probability above reaches a pronounced maximum, i.e. for values of $\xi$ and $k \, a$ satisfying the transcendental equations
\be \label{resonance}
\frac{\cos (2 \gamma k a)}{\cosh (2 \delta  k a)} = - \frac{\delta^2}{\gamma^2} \, , \qquad
\frac{\sin (2 \gamma k a)}{\sinh (2 \delta  k a)} = \frac{\delta^3}{\gamma^3} \, ,
\ee
has already been performed in Ref. \cite{Mostafazadeh:2009zz}, where the high sensitivity to the quantities $\xi$ and $k a$ has been appropriately pointed out. Although the present parametrization is different (and, probably, clearer than in Ref. \cite{Mostafazadeh:2009zz}), we do not further indulge on the theoretical question of the appearance of spectral singularities, but rather focus on the physical effects behind the transmission probability in Eq. (\ref{prob}), which we plot in Fig.~\ref{fig2} both for the ``over-gain" ($E<v$) and ``under-gain" ($E>v$) configurations.
Spectral singularities manifest themselves in the evident probability peaks, which are present in both configurations.

As a general feature, holding irrespective of the value taken by $\xi$ -- which, in a sense, parametrizes the gain/loss contrast of the dielectric --, the transmission probability starts from the unit value for small $k$ (measured in the inverse semi-barrier length unit), while asymptotically tends to zero for large $k$. Between these two extreme cases, in the over-gain configuration ($\xi >1$) just one maximum is present, whose height extends well above the unit value of the transmission probability, for $\xi$ values not very far from 1 (see Fig.~\ref{fig2}, top). This is not surprising in presence of active media, where the unitarity limit does not hold anymore since, in addition to the incident wave, further waves may be emitted by the stimulated medium. For example, by assuming that the active medium of our waveguide is produced by a laser-induced population inversion, as envisaged above, values of the transmission (or reflection) coefficient larger than 1 are caused by the effective conversion of the laser energy into a more intense transmitted (or reflected) wave in the exit region. Note that changing $\xi$ results just into a shift in the position of the peak, as well as a change in the peak height (and width).

In the under-gain configuration, instead, a number of probability peaks is present, whose heights increase with $k$ (for fixed $\xi$) till a maximum (see Fig.~\ref{fig2}, bottom), after which they decrease up to the asymptotic limit value seen above, in a sort of probability wave packet. Intriguingly enough, the different peaks for a given $\xi$ are appreciably {\it equispaced}, and their positions, irrespective of their heights, are the {\it same} for different values of $\xi$ (see the left insert in Fig.~\ref{fig2}, bottom), while the change in the peak heights with different values of $\xi$ is apparently not very regular. Note also that the probability peaks in the under-gain configuration are exceedingly higher than those in the over-gain one, with however a notable exception for $\xi$ around the value $1.82$.

The situation is likely clearer when looking at the contour plots of the transmission probability reported in Fig.~\ref{fig3}. {An extended region centered around $k a \simeq 1$ and $\xi \simeq 2$ exists where the probability reaches values exceedingly higher than 1: this corresponds to the first spectral singularity of the system, with the highest amplitude, but more singularities appear with  increasing $k a$, though with smaller amplitudes. Indeed (see Fig.~\ref{fig3}), such a pattern repeats} itself regularly for $\xi < 1$, the ``resonance" values of $\xi$ decreasing with increasing $k$. The distribution of the probability maxima is shown in the same figure, where the correlation of the ``resonance" values of $\xi$ with $k$ is more evident for both over-gain and under-gain configuration.

\begin{figure*}
\begin{center}
\bt{ll}
\includegraphics[width=12cm]{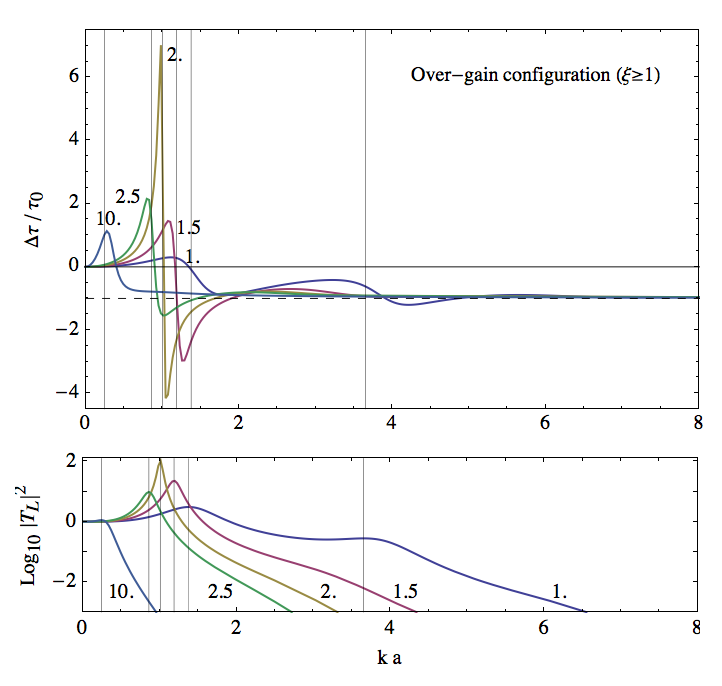}
\et
\caption{Time delay (top) for the complex potential system in Fig.~\ref{fig1} in the over-gain configuration, compared to the transmission probability (bottom). As above, the numbers appearing in each picture report the values of the $\xi$ parameter considered. The horizontal dashed line marks the asymptotic value $\Delta \tau = - \tau_0$ in the $k a \gg 1$ limit, while the vertical lines denote the position (in $k a$) where the maximum transmission probability is attained for the considered value of the $\xi$ parameter.}
\label{fig4}
\end{center}
\end{figure*}

\begin{figure*}
\begin{center}
\bt{ll}
\includegraphics[width=13.5cm]{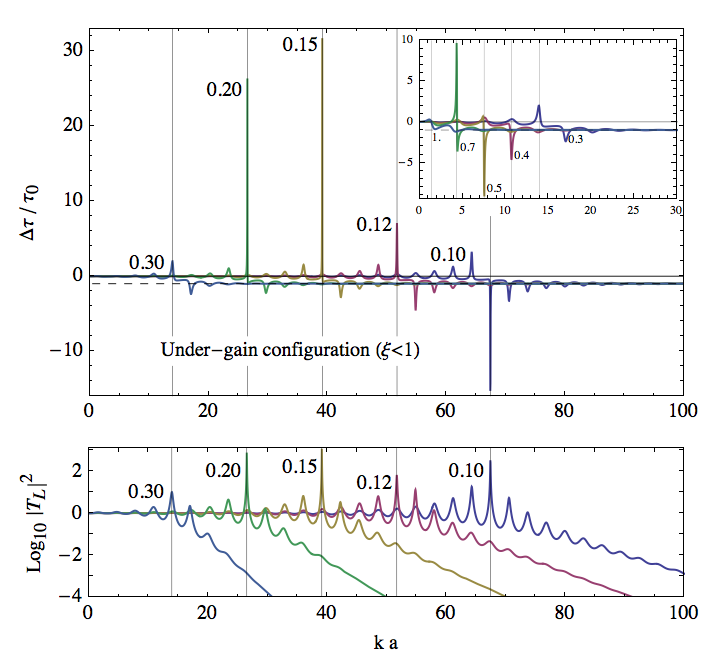}
\et
\caption{The same as in Fig.~\ref{fig4}, but for the under-gain configuration. The vertical lines correspond to the maximum of the probability wave packet.}
\label{fig5}
\end{center}
\end{figure*}

Even more intriguing results come from the calculations of the {experimentally observable (see above)} delay time $\Delta \tau = \tau - \tau_0$ in traversing the potential barrier region ($\tau_0$ is the time needed to the incident wave to traverse the region of length $2 a$ without the barrier), deduced from the phase time in Eq. (\ref{time}):
\be \label{delay}
\Delta \tau = \frac{\tau_0}{2 a} \, \frac{\drm \varphi}{\drm k}  \qquad \left( \tau_0 = 2 a \, \frac{\drm k}{\drm \omega} \right) .
\ee
The quantity $\varphi$ is the phase of the transmission coefficient $T_L$, given by:
\be \label{tan}
\tan \varphi = \frac{\gamma^2 A - \delta^2 B}{\gamma^2 C - \delta^2 D} \, ,
\ee
\bea
& & A = \gamma \sin (2 \gamma k a) \cos (2 k a) - \cos (2 \gamma k a )\sin (2 k a )\, ,
\nonumber \\
& & B = \delta \sinh (2 \delta k a) \cos (2 k a) + \cosh (2 \delta k a) \sin (2 k a) \, ,
\nonumber \\
& & C = \gamma \sin (2 \gamma k a) \sin (2 k a) + \cos (2 \gamma k a) \cos (2 k a) \, ,
\nonumber \\
& & D = \delta \sinh (2 \delta k a) \sin (2 k a) - \cosh (2 \delta k a) \cos (2 k a) \, . \nonumber
\eea
The transit time $\tau$ obviously vanishes in the thin barrier limit ($k a \rightarrow 0$), as can be directly checked from Eq.s (\ref{delay}), (\ref{tan}), since $\Delta \tau \rightarrow 0$, $\tau_0 \rightarrow 0$. However, it surprisingly vanishes even in the opaque barrier limit ($k a \rightarrow \infty$), where $\Delta \tau \rightarrow - \tau_0$ (and this holds independently of the value taken by $\xi$). {In this limit, indeed, the transit time vanishes exponentially (with a sinusoidal modulation) according to the expansion
\be
\tau \simeq  \frac{2 \tau_0}{\delta^2} \, \cos \left( 2 \gamma k a \right) \erm^{- 2 \delta k a} \, + \, \frac{\tau_0}{k a} \, \frac{\delta}{\gamma^2 + \delta^2} \,  ,
\ee
the result being reminiscent of the Hartman effect displayed in conventional barrier tunneling (see, for example, \cite{Esposito:2001qr,Esposito:2002jk}), where the barrier is crossed (with exponentially vanishing probability) in zero phase time.} Between such two limit cases, the behavior of the delay time is not at all trivial, corresponding (partly) to the non-trivial as well behavior of the transmission probability discussed above.

First of all, from Fig.s~\ref{fig4}-\ref{fig5} we note even pronounced {\it negative} delay times in given intervals of $k a$ for any value of the $\xi$ parameter, both for the over-gain and under-gain configuration. It is evidently a manifestation of a sort of non-local effect: the incident wave ``stimulates" the emission of radiation from the active medium even {\it before} it has travelled through the entire system.

In the over-gain configuration, (just) one maximum, positive delay time exists for some value of the parameters, followed by a minimum, negative delay time. Instead, a multi-peak behavior is present in the under-gain configuration (with $\xi =1$ as a limiting case), where the positive (relative) maxima always precede the absolute maximum, while the reverse is true for the minima, although  -- intriguingly -- the greatest absolute delay time is {\it not always} positive (see the cases with $\xi = 0.1, 0.4, 0.5$ in Fig.~\ref{fig5}).

Even more interesting is the result about the relationship between delay time and transmission probability. {Indeed,  although there is apparently no reason to expect maximum delay time  when the transmission probability is as well at a maximum, since the conditions in Eq. (\ref{resonance}) are different from the corresponding ones for maximum delay time,  nevertheless this correspondence becomes approximately true for small $\xi$ values. From Fig.s~\ref{fig4}-\ref{fig5} we can in fact deduce that, in the under-gain configuration, the maximum delay time occurs approximately when the transmission probability takes as well a maximum (a very useful feature in the experimental investigation of the phenomenon), but this property does not hold strictly anymore in the over-gain configuration (with a transition region for yes/no around $\xi =1$). In such configuration, as evident from Fig.~\ref{fig4}, the ``shift" amplitudes (shift between the position of the maximum probability and maximum delay time) are well appreciable, the maximum transmission probability shifting from positive to null or even negative delay times.}

The results presented above, then, clearly show the very rich phenomenology associated to $PT$-symmetric structures. In addition to the unconventional transmission properties studied here, also the scattering properties are expected to reveal intriguing results \cite{Miri}, where the local flow of energy between gain and loss regions induces light to be deflected in unusual ways (depending on the gain/loss contrast), a tilt being induced in the light wavefront while propagating along the gain/loss interface of the considered structure. Despite the impressive experimental efforts in the recent years, a complete phenomenological understanding of such non-trivial systems is nevertheless lacking, and the interesting properties they exhibit, part of which have been discussed here, still expects to be fully explored experimentally.

{The effects of the complex potential considered above can be experimentally studied by letting an electromagnetic wave of frequency $\omega$ to impinge on a rectangular waveguide \cite{Mostafazadeh:2009zz} with height $2b$ and aligned along the $x$-axis, whose regions $-a < x < 0$ and $0 < x < a$ serve as gain and loss regions, respectively, with a plasma frequency $\omega_p$ and damping constant $\delta$. Such a device gives rise to a complex permittivity $\epsilon (x) = 1 - V(x)$, with $v = {\omega_p^{2}}/{2 \delta \omega}$ \cite{Muga}. Maxwell equations for this system reduce to the Schr\"odinger equation for the barrier potential $V(x)$, and the solutions can be identified with transverse electric waves satisfying the appropriate boundary conditions for the waveguide. Spectral singularities manifest provided that, in the given conditions, the experimental parameters are chosen as $\hbar \omega_p \simeq 0.2 eV$, $\hbar \delta \simeq 1.25 eV$. Indeed a spectral singularity could be found for $\hbar \omega = 5 eV$, $a \simeq 1004 nm$ and $b \simeq 62 nm$, as reported in the literature \cite{Mostafazadeh:2009zz}.}

{We expect, however, our results to be valid also for more general potentials, for example potentials with a real part and heterostructures with several layers. These generalizations will be the subject of forthcoming  studies.}


\end{document}